\documentstyle[epsf,preprint,pra,aps]{revtex}

\begin{document}
\preprint{}
\draft

\title{Hybrid quantum computing}

\author{Seth Lloyd${}^{\dagger}$ }
\address{ d'Arbeloff Laboratory for Information Systems and
Technology, \\
Department of Mechanical Engineering,  
Massachusetts Institute of
Technology, \\
Cambridge, Massachusetts 02139 }

\maketitle

\begin{abstract}
Necessary and sufficient conditions are given for the construction
of a hybrid quantum computer that operates on both continuous and
discrete quantum variables.  Such hybrid computers are shown to
be more efficient than conventional quantum computers for performing
a variety of quantum algorithms, such as computing eigenvectors
and eigenvalues.
\end{abstract}



Quantum computers are devices that process information in a way
that preserves quantum coherence 
\cite{feynman82,deutsch85,lloyd93,shor94,lloyd95,divincenzo95,cirac95},
\cite{kimble95,monroe95,pellizzari95}
The most common model of quantum
computation deals with coherent logical operations on two-state
quantum variables known as qubits.  Quantum computation can also
be performed on variables with three or more states, and is well-defined
even when the underlying degrees of freedom are continuous 
\cite{lloydslot98,braunstein98,braunstein98a,lloydbraun99}.  This
paper investigates hybrid quantum computers that operate on both discrete
and continuous quantum variables.  It is shown that a simple set
of operations (hybrid quantum logic gates) can be used to approximate
arbitrary tranformations of the variables.  Hybrid versions of 
quantum algorithms are discussed and a hybrid version of an algorithm
for finding eigenvalues and eigenvectors is presented.  Hybrid
quantum algorithms can have a number of advantages over conventional
quantum algorithms, including lower computational complexity and
an enhanced resistance to noise and decoherence.

The primary reason for investigating hybrid quantum computers is that
nature contains both discrete quantum variables such as nuclear spins,
photon polarizations, and atomic energy levels, and continuous variables
such as position, momentum, and the quadrature amplitudes of the 
electromagnetic field.  In conventional quantum computation, continuous 
variables are something of a nuisance: either they figure as sources
of noise and decoherence, as in the case of environmental baths of
harmonic oscillators, or they must be restricted to a discrete
set of states by cooling, as in the case of the oscillatory modes
of ions in ion-trap quantum computers.  In hybrid quantum computation,
by contrast, the full range of continuous quantum variables can
be put to use.   

The basic model for performing quantum computation using a hybrid of
continuous and discrete variables follows the normal model for performing
quantum computation using discrete or continuous variables on their
own\cite{lloyd95a,lloydbraun99}.  Assume that one has the ability to 
`turn on' and `turn off'
the members of a set Hamiltonian operators $\{ \pm H_j \}$, corresponding
to the ability to apply unitary transformations of the form
$e^{\pm i H_j t}$.  The set of transformations that can be 
constructed in this fashion is the set of transformations of 
the form $e^{-iHt}$ where $H$ is 
a member of the algebra generated from the $H_j$ via commutation:
i.e., since $e^{iH_2 t} e^{iH_1 t}e^{-iH_2 t}e^{-iH_1t} =
e^{-[H_1,H_2]t^2} + O(t^3)$, 
the ability to turn on and turn off $\pm H_1$ and $ \pm H_2$ allows
one effectively to turn on and off $H=\pm i[H_1,H_2]$, etc.  
Transformations of the form $e^{-iHt}$ for non-inifinitesimal $t$
can then be built up from infinitesimal transformations to 
any desired degree of accuracy.

For the sake of ease of exposition, concentrate here on discrete
variables (qubits) that are spins, characterized by the usual
Pauli operators $\sigma_x, \sigma_y, \sigma_z$ and to continuous
variables (qunats) that are harmonic oscillators 
characterized by the usual annihilation and creation operators 
$a, a^\dagger$ ($[a, a^\dagger] = 1$),  and by the `position' and 
`momentum' operators $X=(a + a^\dagger)/2$, $P=(a-a^\dagger)/2i$,
($[X,P]=i$).  It is convenient to think of the harmonic oscillators
as modes of the electromagnetic field with $X$ and $P$ proportional
to the quadrature amplitudes of the mode.  The generalization to
discrete variables with more than two states and to other forms
of continuous variable is straightforward and will be discussed
below.

To perform quantum computations one must be able to prepare one's
variables in a desired state, perform quantum logic operations,
and read out the results.  Assume that it is possible to prepare
the discrete variables in the state $|0\rangle \equiv |\uparrow\rangle_z$,
and the continuous variables in the vacuum state $|0\rangle$: 
$a|0\rangle = 0$.  Assume that it is possible to measure $\sigma_z$
for the discrete variables and $X$ for the continuous variables. 

Now look at performing transformations of the variables.  Begin
with just a pair --- one spin and one oscillator.  Suppose that
one can turn on and turn off the Hamiltonians 
\begin{equation}
\{ \pm \sigma_x X, \pm \sigma_z X, \pm \sigma_z P \} \,,
\label{set}
\end{equation}
As will now be seen, the ability to turn on and off Hamiltonians from this 
set allows one to enact Hamiltonians that are arbitrary polynomials
of the $\sigma$'s, $X$ and $P$.  Note that these Hamiltonians 
all represent interactions between qubits and oscillators: this
is physically realistic in the sense that transformations on physical spins
or atoms are accomplished by making the spins interact with the 
electromagnetic field, and {\sl vice versa}.  In physically realizable
situations, such as the ion traps and optical cavities discussed below,
the interactions in \ref{set} are turned on and off by applying laser
or microwave pulses to couple discrete to continuous degrees of freedom.  

Now investigate what can be accomplished by turning on and off these
interactions.  If the spin is prepared in the state $|0\rangle$,
then turning on the Hamiltonian $\sigma_z P$ is equivalent to
turning on the Hamiltonian $P$ for the oscillator on its own.
The Hamiltonian $X$ can be turned on in a similar fashion.
In order to apply this Hamiltonian for a finite amount of time,
the spin must be constantly reprepared in the state $|0\rangle$
or new spins in this state must be supplied.  This operation
allows the construction of coherent states of the oscillator.

Now start constructing effective Hamiltonians by the method
of commutation above.  Since $i[P,\sigma_x X]=\sigma_x$, we can 
effectively turn on the Hamiltonian $\sigma_x$.  Similarly for
the Hamiltonian $\pm i[P, \sigma_z X] = \pm \sigma_z$.  
And since $i[\sigma_z,\sigma_x] = 2\sigma_y$, any single qubit transformation
$ e^{-i\sigma t} \in SU(2)$ can be enacted by turning on and off Hamiltonians
in the set.  Since $i[\sigma_z P, \sigma_z X] =  2$, an arbitrary
overall phase can also be turned on and off.
That is, we can enact arbitrary single qubit transformations.

Now systematically build up higher order transformations.
Since $i[\sigma_z X, \sigma_x X] = 2\sigma_y X^2$, and
 $i[\sigma_y X^2, \sigma_x X] = 2\sigma_z X^3$, etc.,
we can effectively turn on and off Hamiltonians of the
form $\sigma X^n$, for arbitrary $\sigma$, $n$.  Similarly,
we can turn on and off Hamiltonians of the form $\sigma P^n$.
By preparing the spin in the state $|0\rangle$ and turning on
and off the Hamiltonians $\sigma_z X^m$, $\sigma_z P^n$, we can enact
single oscillator transformations corresponding to Hamiltonians that
are arbitrary Hermitian polynomials in $X$ and $P$.  (Not all such Hamiltonians
are bounded.  Nonetheless, one can build up infinitesimal versions
of such Hamiltonians and apply them for finite time to states 
for which they are bounded.)

So the simple set of Hamiltonians above allows the construction
of arbitrary single qubit transformations and arbitrary polynomial
transformations of the continuous variable, along with arbitrary
interactions between the spin and the oscillator.  Let us now
look at more than one spin and one oscillator.

Since $i[\sigma_z^1 P, \sigma_z^2 X] = \sigma_z^1\sigma_z^2$,
we can turn on the interaction Hamiltonian $\sigma_z^1\sigma_z^2$
between two spins $1$ and $2$ by making them both interact with the
same oscillator.  But the ability to turn on this
Hamiltonian together with the ability to turn on arbitrary single-spin
Hamiltonian translates into the ability to perform arbitrary transformations
on sets of spins: that is, one can perform arbitrary quantum logic
operations on the qubits alone.  

Similarly, since
$i[\sigma_y X_1, \sigma_x X_2] = 2\sigma_z X_1 X_2$, the ability to
make two oscillators interact with the same spin, initially in the
state $|0\rangle$, allows one to turn on the Hamiltonian $X_1X_2$
between the two oscillators $1$ and $2$.  But this ability, together
with the ability to turn on single oscillator Hamiltonians that
are arbitrary Hermitian polynomials in $X$ and $P$, translates into
the ability to turn on Hamiltonians that are arbitrary Hermitian
polynomials of $X_i,P_i$ for all the oscillators together.
So one can perform universal quantum computation on the continuous
variables on their own.

Continuing with constructing Hamiltonians via commutation, 
the ability to prepare the $|0\rangle$ states for spins
and oscillators, together with the ability to turn on and
off the simple set \ref{set} of Hamiltonians given above, allows one
to effectively turn on and off Hamiltonians that are
arbitrary Hermitian polynomials in $ 1, \sigma^j_x,\sigma^j_y, \sigma^j_z,
X^m_k, P^n_k$.  That is, one can perform universal quantum
computation on the hybrid quantum computer.

How might such a hybrid quantum computer be realized?  As it turns
out, many existing designs for quantum computers are easily modified
to perform hybrid quantum computation.  For example, ion trap quantum
computers \cite{cirac95,monroe95}
operate by coupling together the internal states of ions
in an ion trap (qubits) via their motional state (harmonic oscillators).
Existing schemes for performing quantum computation using ion traps
only use the ground and first excited state of the oscillator corresponding
to the fundamental mode of the ions in the trap, 
effectively treating the oscillator as a qubit.  But the same
methods that are used to couple the ions to the oscillator can
just as well be used to apply the Hamiltonians in the set \ref{set}
above.  An ion trap with many ions has many modes, each of which can be used
as a continuous variable in the hybrid quantum computation.   
Similarly, the Pellizzari scheme for coupling together
trapped atoms (qubits) via a cavity mode of the electromagnetic
field can readily be altered to use the quadrature amplitudes of
the modes of the cavity, rather than simply using the lowest two
energy eigenstates of a mode as a qubit \cite{pellizzari95}.  
Other potential continuous
variables that might be used for hybrid quantum computation are the 
translational states of atoms in a Bose condensate, the continuum states 
of electrons in semiconductors, or the state of a Josephson junction circuit.
Essentially any hybrid system that affords precise control over
the interactions between discrete and continuous variables is a good
candidate for a hybrid quantum computer.

An important concern in the construction of hybrid quantum computers
is the problem of noise and decoherence.  At first it might seem
that continuous variables are likely to be more susceptible to noise
than discrete variables.  It is indeed true that more things can go 
wrong with a continuous variable than with a discrete variable.  However,
quantum error correction routines for continuous variables have been
developed and require no greater overhead than those for discrete
variables\cite{lloydslot98,braunstein98,braunstein98a,preskill00}.  
Although these routines are not yet technologically
practical on existing devices, it may well be that improved versions
of these routines combined with existing discrete quantum error correction
routines will allow efficient quantum error correction for hybrid
devices.  In addition, as noted above, hybrid devices have the advantage
that they include in the computation states and degrees of freedom that would
normally be sources of noise, decoherence, and loss.

Now turn to applications of hybrid quantum computers.  Where does
the ability to perform manipulations of continuous variables as well
as qubits give an advantage?   The first point to note in constructing
hybrid algorithms is that we must be careful to assume physically
reasonable uses of hybrid variables---i.e., uses that do not require
infinite or exponentially high precision.  Even in the classical case,
the use of continuous variables can give remarkable computational
speed ups (the ability to solve NP-complete problems in polynomial time, 
the ability to find the the answer to uncomputable problems in finite
time, etc.) if one allows arbitrary precision in manipulating and 
measuring continuous variables.  By giving an explicit construction 
of the operations that can be used to perform continuous variable
and hybrid quantum computation, however, we have implicitly avoided
the use of infinite or excessive precision: all such operations would
require infinite or excessive computational resources to construct,
manipulate, and measure the desired over-precise states. 

With this caveat in mind, turn to the operations that are relatively
easy to perform using continuous quantum variables.  A particularly
useful subroutine in a variety of quantum algorithms is the quantum
Fourier transform: $|x\rangle \rightarrow \sum_{y=1}^q e^{ixy} |y\rangle$.
In the case of discrete quantum variables
the quantum Fourier transform on $N$ qubits takes on
the order of $N$ quantum logic operations to perform.
Although this is an efficient algorithm it is nonetheless
difficult at present to perform quantum Fourier transforms
on more than a few qubits (the current record is three)\cite{weinstein99}.
By contrast, in the case of the continuous quantum variables
$X$ and $P$, the quantum Fourier transform is trivial.
If the eigenstates of $X$ with eigenvalue $x$ are written
$|x\rangle$, then the eigenstates of $P$ with eigenvalue
$p$ can be written $|p\rangle = (1/\sqrt{2\pi}) \int_{-\infty}^{\infty}
e^{ipx}|x\rangle dx$.  That is, the eigenstates of $P$ are
the quantum Fourier transform of the eigenstates of $X$.
Applying the Hamiltonian $X^2+P^2$ for a period of time $\pi/2$
takes $X\rightarrow P$ and performs the Fourier transform.
The quantum Fourier transform on a continuous variable
is accomplished by a single-step operation.
The ease of performing the quantum Fourier transform on continuous
variables suggests that in devising algorithms for hybrid quantum
computers we look for problems in which the quantum Fourier transform
plays a central role.

Perhaps the best known quantum algorithm in which the quantum
Fourier transform plays a central role is Shor's algorithm for factoring 
large numbers\cite{shor94}.  Setting aside the difficulty of performing
the other operations in this algorithm (such as modular exponentiation),
it is immediately clear that using a continuous variable as the register
on which to perform the quantum Fourier transform in Shor's algorithm
would require an exponentially high precision in the preparation
and manipulation of the continuous variable.  (Hybrid quantum computation
might still be used to speed up some aspects of Shor's algorithm;
this possibility will be investigated elsewhere.)  

A second problem in which the quantum Fourier transform plays
a key role is that of simulating the dynamics of quantum
systems \cite{feynman82,lloyd96,wiesner96,zalka98,abramslloyd99}. 
Comparison with \cite{lloyd96} shows that the ability of hybrid quantum 
computers to turn on and off simple Hamiltonians involving a few discrete
and a few continuous variables at a time translates into the
ability to perform efficient quantum simulations of hybrid
systems.
 
A particularly valuable type of quantum simulation is one that
allows the computation of spectra: using
methods developed in \cite{kitaev95,cleve98,zalka98} Abrams and
Lloyd have developed algorithms for computing eigenvalues and
eigenvectors of quantum systems and for obtaining improved estimates
of the ground state\cite{abramslloyd99}.  In its original discrete form, the 
algorithm is somewhat involved.  However, the fact that quantum Fourier
transforms are straightforward to perform on continuous variables makes
the Abrams-Lloyd algorithm 
particularly simple in the case of hybrid quantum computation.
Here we show how to perform a quantum computation that computes the
eigenvectors of a hybrid system and that writes the eigenvalues
of the system onto a register consisting of a single continuous
variable.  The algorithm is a hybrid version of the
discrete algorithms proposed in \cite{kitaev95,cleve98,zalka98,abramslloyd99}
and is closest in form to the discrete algorithm proposed in \cite{zalka98}
for simulating von Neumann measurements on a quantum computer.    
Independently, Travaglione and Milburn \cite{travaglione00}
have shown how methods of hybrid quantum computation can be used to
compute the eigenvectors of a continuous system and write the eigenvalues
onto a discrete register.

First, prepare a single continuous
variable such as a mode of the electromagnetic field
in the squeezed state $|x=0\rangle =
(1/\sqrt{2\pi}) \int_{-\infty}^{\infty} |p\rangle dp$.
In any practical experiment, of course, such perfectly
squeezed states are unavailable.  Imperfectly squeezed
or unsqueezed states will also work, however.  As discussed below, the
effect of imperfect squeezing is to decrease the resolution
to which the spectrum can be obtained.
Prepare a second system in the state $|\psi\rangle$ whose
decomposition into energy eigenstates
$|\psi \rangle = \sum_i \psi_i|E_i\rangle$ one wishes to obtain.
Here we assume that the system is discrete; in general,
however, system may be continuous, discrete, or a hybrid
of continuous and discrete variables.

Next, using the methods of hybrid quantum computation described
above, couple the system to the continuous variable 
via the coupling Hamiltonian $HP$,
where $H$ is the Hamiltonian whose eigenvalues and
eigenvectors are to be obtained.  For $H$ to be
efficiently simulatable, it must be equal to
$\sum_k H_k$, where each $H_k$ acts on only a few
variables at a time.  Since $HP= \sum_k H_kP$,
if $H$ is efficiently simulatable, so is $HP$, by the methods
of hybrid quantum computation described above.
Writing $H=\sum_j E_j |E_j\rangle\langle E_j|$, the
time evolution of the state of the system and the continuous
variable is 
\begin{eqnarray}
& |\psi\rangle|x=0\rangle \nonumber \\
\rightarrow 
& e^{-iHPt} |\psi\rangle|x=0\rangle \nonumber \\
=
& e^{-i\sum_j E_j |E_j\rangle \langle E_j| Pt} 
\sum_j \psi_j |E_j\rangle|x=0\rangle \nonumber \\
= 
& \sum_j e^{-iE_jtP}\psi_j |E_j\rangle|x=0\rangle \nonumber \\
=
& \sum_j \psi_i |E_j\rangle |x=E_jt\rangle \,,
\label{eigen}
\end{eqnarray}
since $e^{-iPt}|x\rangle = |x+t\rangle$.
Clearly, at this point, a measurement of the variable $X$ on
the continuous variable will yield the result
$x=tE_i$ with probability $\psi_i$, leaving the system
in the state $|E_i=x/t\rangle$.  That is, one can sample
the spectral decomposition of $|\psi\rangle$, obtaining
the eigenvalues $E_i$ together with their corresponding
weights $|\psi_i|^2$ and eigenvectors $|E_i\rangle$.
The process is highly efficient, requiring only the ability
to prepare the initial squeezed state $|x=0\rangle$
and to apply the Hamiltonian $HP$.  

The hybrid eigenvalue and eigenvector finding algorithm using a continuous
variable to register the eigenvalue is more efficient than the
corresponding algorithm using qubits to register the eigenvalue.
Since the quantum Fourier transform is performed implicitly
in the continuous register, fewer steps are required in the
hybrid algorithm.  In addition, unlike the conventional version
of the algorithm, the hybrid version is insensitive to approximate 
decoherence of the register in the course of the computation:
measuring the value $x$ of the register in the course of the 
coupling does not affect the ability of the algorithm to find
eigenvectors and eigenvalues. 

The requirement that the initial state of the continuous
variable be perfectly squeezed can also be relaxed.  Suppose that the
initial state is in a Gaussian state $\int e^{-\beta x^2/2}
|x\rangle dx$.  For example, $\beta=1$ gives the unsqueezed
$n=0$ vacuum state, while $\beta >1 $ gives partial squeezing
in $X$.   With this initial state for the continuous variable,
after the algorithm has been run, the continuous
variable and the system are in the state
\begin{equation}
\sum_j\int e^{-\beta x^2/2} |E_j\rangle|x+E_jt\rangle dx\,.
\label{squeeze}
\end{equation}
That is, the eigenvalues and eigenvectors are resolved
to within an accuracy $1/t\sqrt\beta$.  By coupling the
system to the continuous variable for a sufficiently
long time, the eigenvectors and eigenvalues of $H$ may
be determined to an arbitrary degree of accuracy, even
when the initial state is unsqueezed.  Note that resolving
the eigenvalues of a system with an exponentially large
number of states requires exponential squeezing of the
pointer state.  But as noted in \cite{abramslloyd99},
this algorithm still provides a potentially exponential
speedup over classical algorithms even when the eigenvalues
are not determined to an exponential degree of accuracy.

Hybrid quantum computers are devices that perform quantum computations
using both discrete variables such as spins and continuous variable
such as position and momentum, or the quadrature amplitudes of
the electromagnetic field.  Hybrid quantum computation represents
a natural extension of quantum computation using quantum bits alone:
as the example of finding eigenvalues and eigenvectors presented here shows,
hybrid quantum computations can be more efficient and less sensitive
to noise and decoherence than conventional quantum computations.
Nature supplies us with both discrete and continuous quantum variables:
it is advantageous to use them.

\vspace*{5mm}

This work was supported by DARPA/ARO under the QUIC initiative. \\
${}^\dagger$ {\tt slloyd@mit.edu}


\end{document}